\documentclass[runningheads]{llncs}
\usepackage[T1]{fontenc}
\usepackage{graphicx}
\usepackage{url}
\usepackage{glossaries}
\usepackage{hyperref}

\newacronym[plural=nns, firstplural= Neural Networks (NNs)]{nn}{NN}{Neural Network}
\newacronym{lidar}{LiDAR}{Light Detection And Ranging}
\newacronym{blue}{BLUE}{roBot for Localization in Unstructured Environments}
\newacronym{yolo}{YOLO}{You Only Look Once}
\newacronym{sota}{SOTA}{State Of The Art}
\newacronym{apas}{APAS}{Average Precision Across Scale}
\newacronym[plural=sris, firstplural= Spherical Range Images (SRIs)]{sri}{SRI}{Spherical Range Image}
\newacronym{bb}{BB}{Bounding Box}
\newacronym{cpu}{CPU}{Central Processing Unit}
\newacronym{gpu}{GPU}{Graphics Processing Unit}
\newacronym{ram}{RAM}{Random-Access Memory}
\newacronym{fps}{FPS}{Frames Per Second}
\newacronym{ai}{AI}{Artificial Intelligence}
\newacronym{ssd}{SSD}{Single Shot Detector}
\newacronym{fov}{FOV}{Field Of View}

\usepackage{color}

\begin{document}
\title{LiCAR: pseudo-RGB LiDAR image for CAR segmentation}
%
%
\author{Ignacio de Loyola Páez-Ubieta\orcidID{0000-0001-9901-7264}, Edison P. Velasco-Sánchez\orcidID{0000-0003-2837-2001} \and Santiago T. Puente\orcidID{0000-0002-6175-600X}}
\authorrunning{I.de L. Páez-Ubieta et al.}
%
\institute{AUtomatics, RObotics, and Artificial Vision Lab, IUII: University Institute for Computer Research University of Alicante, Crta. San Vicente s/n, San Vicente del Raspeig, E-03690, Alicante, Spain \email{\{edison.velasco, ignacio.paez, santiago.puente\}@ua.es}}

\maketitle              
\begin{abstract} With the advancement of computing resources, an increasing number of Neural Networks (NNs) are appearing for image detection and segmentation appear. However, these methods usually accept as input a RGB 2D image. On the other side, Light Detection And Ranging (LiDAR) sensors with many layers provide images that are similar to those obtained from a traditional low resolution RGB camera. Following this principle, a new dataset for segmenting cars in pseudo-RGB images has been generated. This dataset combines the information given by the LiDAR sensor into a Spherical Range Image (SRI), concretely the reflectivity, near infrared and signal intensity 2D images. These images are then fed into instance segmentation NNs. These NNs segment the cars that appear in these images, having as result a Bounding Box (BB) and mask precision of 88\% and 81.5\% respectively with You Only Look Once (YOLO)-v8 large. By using this segmentation NN, some trackers have been applied so as to follow each car segmented instance along a video feed, having great performance in real world experiments.

\end{abstract}

\section{Introduction}
\label{sec:introduction}
The first \gls{ai} models date back to 1943 \cite{1943logical}. Initially, the goal was to simulate the function of a single neuron. Since then, several advances have led to the development of more complex structures, the most recent of which are data storage \cite{2006fast} and \glspl{gpu} \cite{2010deep}, in 2006 and 2010 respectively.

Regarding object detection and segmentation methods, there are three different approaches; single, two and multi-stage methods \cite{2022review}. Among them, the first one is the one that receives more attention, since they are usually faster \cite{fast2020review}, allowing to have inference times close to realtime. Some examples of these single-stage methods are \gls{yolo} or \gls{ssd} \cite{ssd2016}.

Normally, these methods require a RGB image as input. However, variations have been introduced in recent years that allow the input of pseudo-RGB images generated from \gls{lidar} sensors.  
For example, \cite{lidarRGBDetect2021} uses a combination of RGB and \gls{lidar} images to perform object detection by first identifying regions of interest in the \gls{lidar} point cloud. Using these regions, an RGB image is extracted from the original RGB image and used as input to a \gls{yolo} \gls{nn}, which finally detects the cars. Although this method gives good results, it requires the handling of \gls{lidar} and RGB images and does not detect objects accurately.
Other papers, such as \cite{lidarDetect2023}, generate a pseudo-RGB image from a \gls{lidar} sensor. This is used to detect and track pedestrians across the image. To do this, they slide the \gls{bb} in 5 parts to build the descriptor. However, it only performs object detection, not segmentation, using an old \gls{yolo} distribution, since at the time of submission new \gls{yolo} distributions had appeared.
Finally, \cite{lidarSegment2023} generates a pseudo-RGB image from a \gls{lidar} sensor in a similar way. Using detection and segmentation methods, they achieve good results in both indoor and outdoor environments. However, the need to pre-process the image from 2048*128 to 1000*300 and the small number of images and instances in their dataset question the quality of the results obtained.

In this work, we perform the segmentation of car instances on \glspl{sri}, obtained by combining three different channels from a \gls{lidar} sensor, with several single-stage methods from the \gls{yolo} family.  

The main contributions of this work are: 
\begin{itemize}
  \item A new dataset for segmenting cars in pseudo-RGB 2D images has been generated. These images are \gls{sri} obtained by combining several channels from a \gls{lidar} sensor. 
  \item Several \gls{sota} \gls{nn} instance segmentation methods have been trained on the aforementioned dataset, achieving almost 90\% in \gls{bb} detection and more than 80\% in mask segmentation.
  \item Once the best instance segmentation was obtained, a tracker was incorporated, achieving excellent results without affecting real-time inference.
\end{itemize}

This paper is organised as follows: Section \ref{sec:methodology} introduces both the image generation process and the segmentation \glspl{nn} used, Section \ref{sec:results} shows the experimental setup of the system, Section \ref{subsec:results} shows the results obtained on the test set and also on completely new real world scenarios and Section \ref{sec:conclusion} summarises the results of the paper, as well as introduces the main lines of future works.

\section{Methodology}
\label{sec:methodology}
In this Section, the process used to generate the images and the segmentation \glspl{nn} that were used in Sec. \ref{sec:results} are presented.

\subsection{Image generation}
\label{subsec:image_generation}
Generating images from a \gls{lidar} sensor involves two steps. In the first one, a pseudo-RGB pointcloud is generated from the data provided by the \gls{lidar}, and in the second one, the previously generated pointcloud is transformed into a \gls{sri}. 

For the first process, data is obtained from the \gls{lidar} sensor. The used sensor provides 4 different 2D images: near infrared photons, range information, calculated calibrated reflectivity and signal intensity photons images. From them, a pseudo-RGB point cloud is generated by combining the calculated calibrated reflectance, NIR photon and signal intensity photon images. 

For the second one, the generated pointcloud is transformed into a \gls{sri}. Since the used \gls{lidar} has enough resolution, no interpolation or filtering process is needed to generate the final \gls{sri} \cite{lilo}.

\begin{figure}
\includegraphics[width=\textwidth]{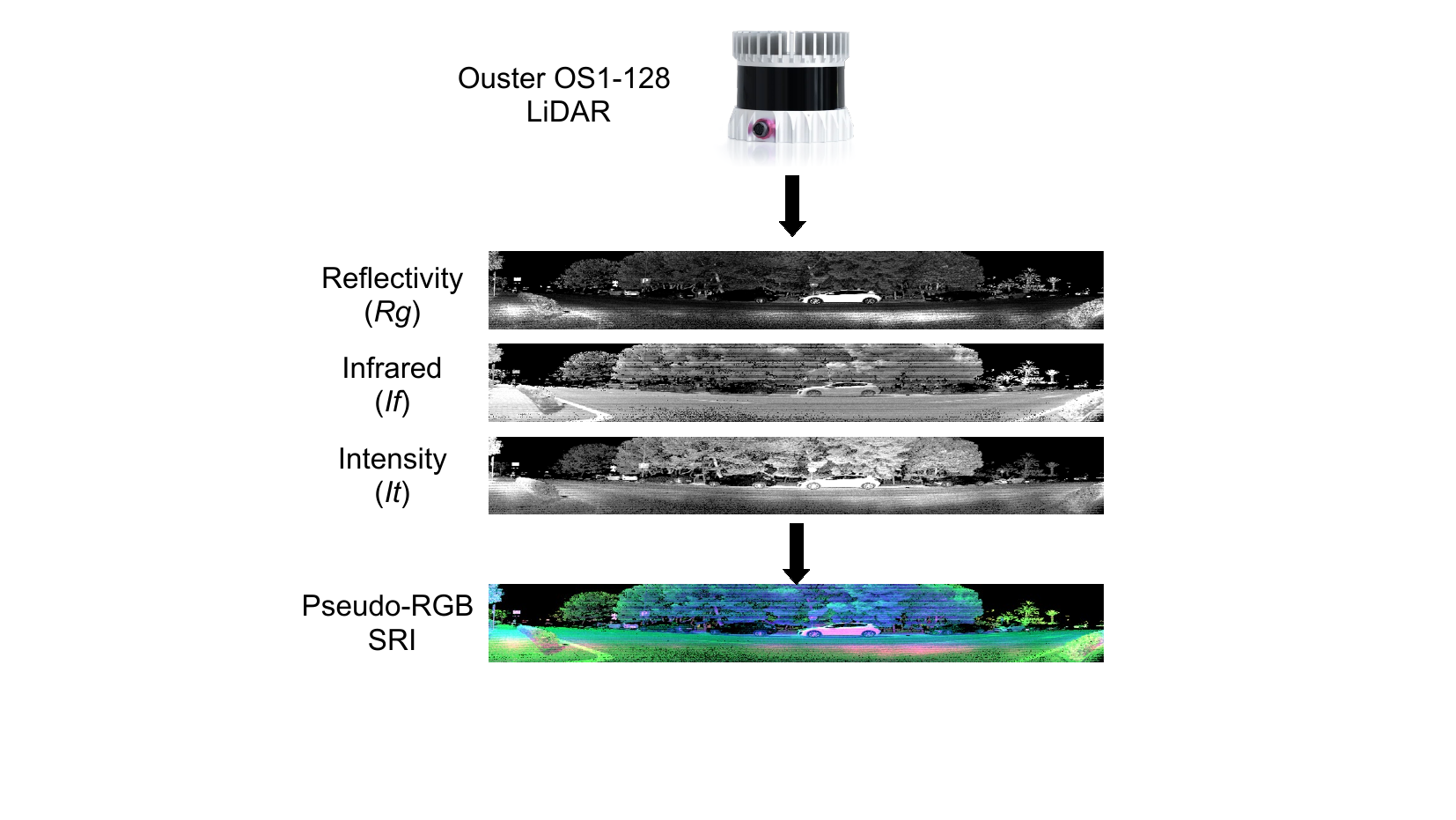}
\caption{Combination of LiDAR channels for generating the pseudo-RGB image.} 
\label{fig:lidar_combination}
\end{figure}

An example of the process of combining the \gls{lidar} channels to produce the final pseudo-RGB \gls{sri} is shown on Fig. \ref{fig:lidar_combination}.

\subsection{Segmentation NNs}
\label{subsec:segmentation_nn}
A fast but accurate method is required to perform instance segmentation. For this purpose, single-stage \gls{sota} methods such as \gls{yolo}-v5 \cite{yolov5}, \gls{yolo}-v7 \cite{yolov7} and \gls{yolo}-v8 \cite{yolov8} were chosen.

\gls{yolo}-v5 appeared online on 2020. This model was the first in the \gls{yolo} \glspl{nn} family to support instance segmentation, as well as being released by a private company. There is not much hindsight on this model, as no official paper was published. But it seems that the anchor box selection process was built into the model, allowing it to automatically learn the most appropriate size for each dataset \cite{yolov5_explain}.

\gls{yolo}-v7 appeared online on 2022, but the paper was published in 2023. They introduced a trainable bag-of-freebies, a new reparametrization module, and a composite scaling method. All these features were used to achieve an improvement in the loss function, label assignment, support for multiple architectures, or a good trade-off between network parameters, computation, inference speed, and accuracy.

\gls{yolo}-v8 also appeared online on 2022, but later than \gls{yolo}-v7, having its paper publication in 2024. To improve on previous versions, a new loss function called focal loss, a new data enhancement method called mixup \cite{mixup}, and a new evaluation metric called \gls{apas} were introduced. These allow to focus on complex images instead of simple ones, to improve the generalisation and strength of the model by merging images and labels, and to ensure the accuracy of the model over different scales.

\section{Experiments}
\label{sec:results}
In this Section, the elements used for the execution of the experiments are presented, as well as details generated dataset. The \glspl{nn} training procedure and the results obtained are shown in the next section.

\subsection{Setup}
\label{subsec:setup}
\begin{figure}
\centering
\includegraphics[width=0.5\textwidth]{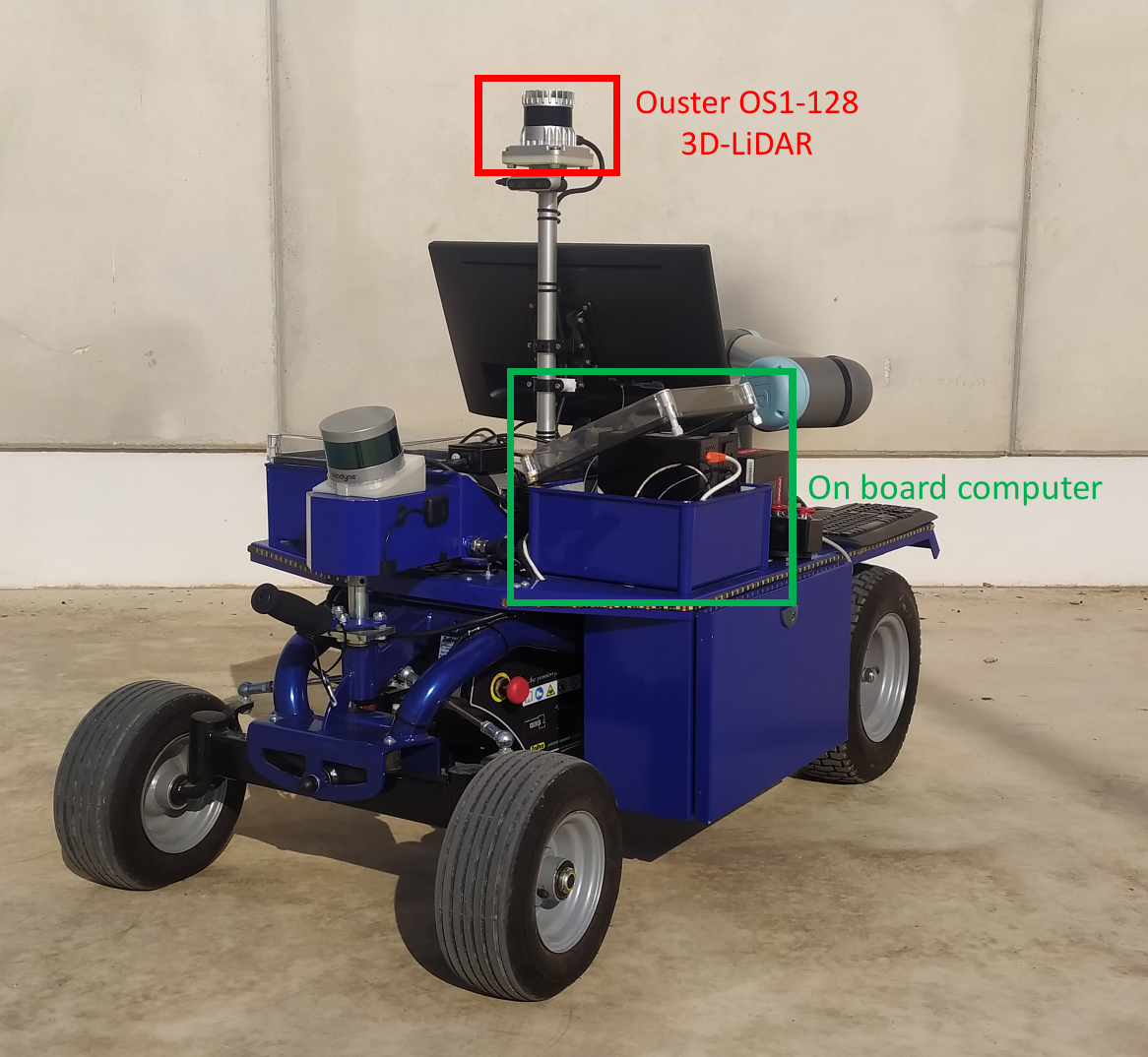}
\caption{\gls{blue} robot with both the Ouster OS1-128 3D-\gls{lidar} and the on board computer.} 
\label{fig:blue}
\end{figure}

The Ouster OS1-128 3D-\gls{lidar} was used to obtain the dataset. This sensor is mounted on the \gls{blue} robot, a mobile robotic platform designed by the research group several years ago \cite{presenting_blue,deeper_blue}, but which is still being updated today \cite{desarrollos_blue}. The \glspl{nn} was trained on a computer with a NVIDIA DGX-A100 Tensor Core GPU with 40GB of RAM. For data acquisition and inference, an on board computer with an AMD Ryzen 7 5700G as \gls{cpu}, a NVIDIA RTX 3060 with 12 GB as \gls{gpu} and 32 GB of \gls{ram} was used. The \gls{blue} robot with Ouster OS1-128 3D-\gls{lidar} and on board computer is shown in Fig. \ref{fig:blue}.

\subsection{Dataset}
\label{subsec:dataset}
After obtaining the data from the \gls{lidar} sensor and converting it to a \gls{sri}, 400 images were obtained, each with approximately 14 car instances. These images contain different outdoor environments from the University of Alicante and have a resolution of 2048x128 pixels. The cars appearing in the images were labelled using the LabelMe \cite{labelme} tool. These generated labels include a special encoding used to generate \gls{yolo} \glspl{nn} to understand that a partially occluded object belongs to the same instance, so that it is included in the same \gls{bb} and mask instance.

The generated dataset has been divided into training/validation/test sets according to the 85/10/5 ratio. A summary of the generated dataset can be found in Table \ref{tab:dataset}. This dataset is available online at \href{https://www.iuii.ua.es/licar/index.html}{LICAR dataset}.

\begin{table}
\caption{Generated dataset specifications.}
\label{tab:dataset}
\centering
\begin{tabular}{c|c|c}
\hline
Set        & Images & Instances \\ \hline\hline
Train      & 340    & 4784      \\ \hline
Validation & 40     & 569       \\ \hline
Test       & 20     & 299       \\ \hline
\end{tabular}
\end{table}

\subsection{NN training}
\label{subsec:nn_training}
All trainings were performed using transfer learning from previously trained models provided by each \gls{nn} model, as it has shown to improve model accuracy while training the models in \cite{transferLearning}.

\gls{yolo}-v5 was trained for 300 epochs with a batch size of 128 and 8 workers, using a single channel rectangular image with an image size of 2048 pixels as the longest side. \gls{yolo}-v7 was trained for 100 epochs with a batch size of 64 and only 1 worker, also using single channel rectangular images of 2048 pixels. \gls{yolo}-v8 was trained for 100 epochs with a batch size of 256, using again the single channel rectangular 2048 pixel rectangular image.

The rest of the used parameters used in each case are the default ones provided by each of the models as recommended parameters for performing training of instance segmentation models.

\section{Results}
\label{subsec:results}

This section presents the results of the experiments described on Section \ref{sec:results}.

\subsection{Test set}
\label{subsubsec:testSet}

After training the aforementioned models using the training and validation sets, the final results are obtained. The test set contained 20 images with 299 instances, representing all types of data used during the training process. However, in order to have a fair comparison, some parameters of the used \glspl{nn} are given in Table \ref{tab:nn_specifications}, as well as the computation times per image (executed in the onboard BLUE computer with a NVIDIA RTX 3060 \gls{gpu}). These computation times are expressed as the sum of three values, which consist of the time spent during preprocess, inference and postprocess steps. So, for example, \gls{yolo}-v7 has a similar number of layers to \gls{yolo}-v8 medium, but is between \gls{yolo}-v8 medium and \gls{yolo}-v8 large in terms of the number of parameters. Another example is \gls{yolo}-v5, which has fewer layers than \gls{yolo}-v8 nano, but is between \gls{yolo}-v8 small and \gls{yolo}-v8 medium in terms of parameters.

\begin{table}
\caption{Some NN specifications.}
\label{tab:nn_specifications}
\centering
\begin{tabular}{c|c|c|c}
\hline
\gls{nn}             & Speed (ms)   & Number of layers & Number of parameters \\ \hline\hline
\gls{yolo}-v8 nano   & 0.3+6.5+0.5  & 261              & 3263811              \\ \hline
\gls{yolo}-v8 small  & 0.3+8.3+0.5  & 261              & 11790483             \\ \hline
\gls{yolo}-v8 medium & 0.3+18.2+0.6 & 331              & 27240227             \\ \hline
\gls{yolo}-v8 large  & 0.3+28.7+0.5 & 401              & 45912659             \\ \hline
\gls{yolo}-v7        & 0.9+23.7+0.7 & 325              & 37842476             \\ \hline
\gls{yolo}-v5 medium & 0.9+15.9+0.7 & 220              & 21652358             \\ \hline
\end{tabular}
\end{table}

The results for \gls{bb} detection and mask segmentation are shown in Table \ref{tab:bb_results} and \ref{tab:mask_result} respectively. In terms of \gls{bb} detection, \gls{yolo}-v8 medium and large have very close results, with the precision being higher for the \gls{yolo}-v8 large model. However, when moving to the mask segmentation task, \gls{yolo}-v8 would be the model with the best results in 3 out of the 4 metrics evaluated.

Also, all the trained models fall within the real-time specification, as their computation time should be between 33.3 and 66.6 ms for 30 and 15 \gls{fps}. So our final choice is \gls{yolo}-v8 large.

\begin{table}
\caption{Results in BB detection.}
\label{tab:bb_results}
\centering
\begin{tabular}{c|c|c|c|c}
\hline
\gls{nn}             & Precision     & Recall         & mAP @0.5       & mAP@0.5-0.95   \\ \hline\hline
\gls{yolo}-v8 nano   & 0.858         & 0.756          & 0.85           & 0.589          \\ \hline
\gls{yolo}-v8 small  & 0.836         & 0.832          & 0.871          & 0.633          \\ \hline
\gls{yolo}-v8 medium & 0.824         & 0.832          & \textbf{0.888} & \textbf{0.654} \\ \hline
\gls{yolo}-v8 large  & \textbf{0.88} & 0.811          & 0.881          & 0.649          \\ \hline
\gls{yolo}-v7        & 0.81          & 0.823          & 0.87           & 0.62           \\ \hline
\gls{yolo}-v5 medium & 0.783         & \textbf{0.835} & 0.849          & 0.587          \\ \hline
\end{tabular}
\end{table}

\begin{table}
\caption{Results in mask segmentation.}
\label{tab:mask_result}
\centering
\begin{tabular}{c|c|c|c|c}
\hline
\gls{nn}             & Precision      & Recall         & mAP @0.5       & mAP@0.5-0.95   \\ \hline\hline
\gls{yolo}-v8 nano   & 0.795          & 0.676          & 0.758          & 0.448          \\ \hline
\gls{yolo}-v8 small  & 0.77           & 0.749          & 0.803          & 0.471          \\ \hline
\gls{yolo}-v8 medium & 0.782          & 0.732          & 0.807          & 0.477          \\ \hline
\gls{yolo}-v8 large  & 0.815          & \textbf{0.751} & \textbf{0.826} & \textbf{0.509} \\ \hline
\gls{yolo}-v7        & 0.803          & 0.679          & 0.769          & 0.474          \\ \hline
\gls{yolo}-v5 medium & \textbf{0.831} & 0.605          & 0.732          & 0.431          \\ \hline
\end{tabular}
\end{table}

\begin{figure}
\includegraphics[width=\textwidth]{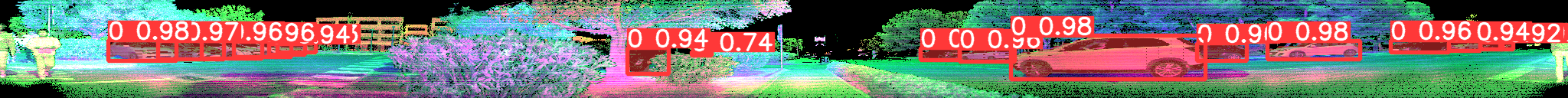}
\includegraphics[width=\textwidth]{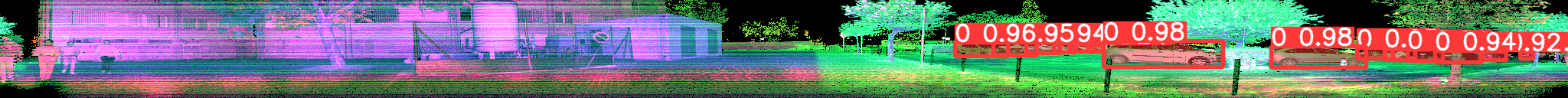}
\includegraphics[width=\textwidth]{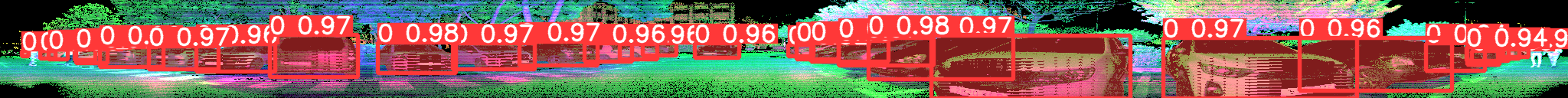}
\caption{Some samples from test set segmented by \gls{yolo}-v8 large.} 
\label{fig:results_yolov8l}
\end{figure}

Some images of the test subset segmented by \gls{yolo}-v8 large can be seen in Fig. \ref{fig:results_yolov8l}. As can be seen, cars are correctly segmented in both near and far distance in all images, representing the 360 degrees of the \gls{lidar} sensor \gls{fov}.

\subsection{Real experiments}
\label{subsubsec:Real experiments}
\begin{figure}
\includegraphics[width=\textwidth]{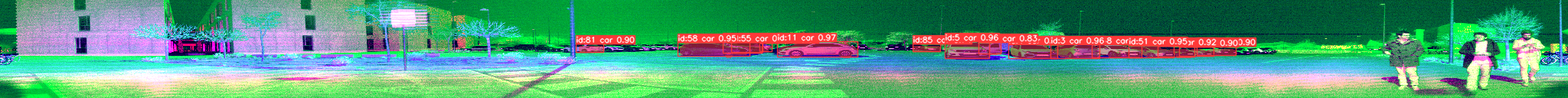}
\includegraphics[width=\textwidth]{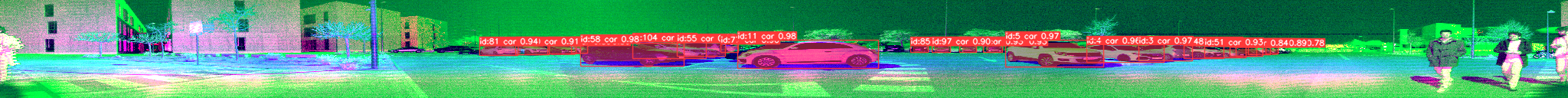}
\includegraphics[width=\textwidth]{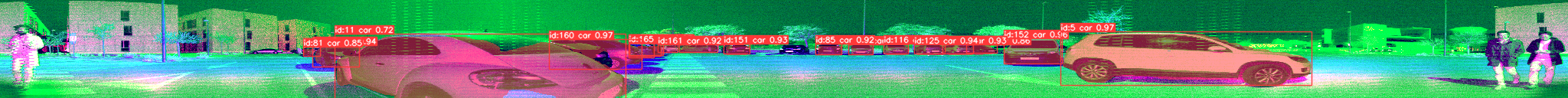}
\caption{Several frames running \gls{yolo}-v8 large using BoT-SORT as tracker.} 
\label{fig:video_frames}
\end{figure}

Once the best segmentation method was found, it was tested in a completely new and unseen environment. Also some trackers like BoT-SORT \cite{botsort} and ByteTrack \cite{bytetrack} were added to keep the position of different segmented instances across frames on a video. Some frames of a \href{https://youtu.be/FNemm0QZXo4}{sample video} using the BoT-SORT tracker are shown in Fig. \ref{fig:video_frames}. This tracker was chosen over ByteTrack because it represents an evolution, including some improvements such as a camera compensation based feature tracker and a suitable Kalman filter.

BoT-SORT was used with 0.7 as the threshold for first and second association, 0.75 as the threshold for initialising a new track if the detection does not match any previous track, 20 as the buffer size for removing tracks and 0.8 as the threshold for matching tracks.

\section{Conclusion}
\label{sec:conclusion}
This paper presents the segmentation of instances in \gls{sri} using several single-step methods such as \gls{yolo}-v5, \gls{yolo}-v6 and \gls{yolo}-v8. Specifically, the aim was to detect cars in pseudo-RGB images obtained from a \gls{lidar} sensor by combining several of the different channels available. The results have shown that it is possible to perform instance segmentation in a \gls{lidar} image using \gls{yolo}-v8 large, having a precision of 81.5\%. This model provided the best accuracy in mask segmentation, allowing real-time inference on \gls{blue}'s on-board computer and the mobile robot to segment cars in outdoor environments. 

Future work will consist of using the designed system to segment different types of objects, as well as designing a new method for calibrating \gls{lidar} and the camera using the segmented masks obtained with this method.

\begin{credits}
\subsubsection{\ackname} Research work was funded by grant PID2021-122685OB-I00 funded by MICIU/AEI/10.13039/501100011033 and ERDF/EU and grand PRE2019-088069 funded by MICIU/AEI/10.13039/501100011033 and ESF Investing in your future. The computer facilities were provided through the IDIFEFER/2020/003 project.

\subsubsection{\discintname}
The authors have no competing interests to declare
that are relevant to the content of this article.

\end{credits}

\bibliographystyle{splncs04}
\bibliography{paper_bib}

\end{document}